\documentclass[pra,letterpaper,aps,10pt,superscriptaddress,twocolumn,floatfix,showpacs]{revtex4-1}
\usepackage{graphicx}
\usepackage{amsmath}
\usepackage{amsfonts}
\usepackage{float}
\usepackage{amssymb}
\usepackage{epsfig}
\usepackage{epstopdf}
\DeclareGraphicsExtensions{.pdf,.eps,.png,.jpg,.mps}
\usepackage[pdftex]{color}
\usepackage{amsmath,graphicx,amssymb,braket,xcolor,subfigure,upgreek}
\usepackage[colorlinks, linkcolor=blue, citecolor=blue, urlcolor=blue, breaklinks=true]{hyperref}
\usepackage{microtype}
\usepackage{bbm}
\usepackage{color}
\usepackage{dsfont}
\usepackage{xcolor}
\usepackage{wrapfig}

\begin{document}


\title{{Control of Localized Multiple Excitation Dark States in Waveguide QED}}


\author{R. Holzinger}
\affiliation{Institute for Theoretical Physics, Innsbruck University, Technikerstrasse 21a, 6020 Innsbruck, Austria}

\author{R. Guti\'{e}rrez-J\'{a}uregui}
\affiliation{Department of Physics, Columbia University, New York, NY 10027, USA}

\author{T. H\"{o}nigl-Decrinis}
\affiliation{Institute for Quantum Optics and Quantum Information of the Austrian Academy of Sciences,
6020 Innsbruck, Austria}

\author{G. Kirchmair}
\affiliation{Institute for Quantum Optics and Quantum Information of the Austrian Academy of Sciences,
6020 Innsbruck, Austria}

\author{A. Asenjo-Garcia}
\affiliation{Department of Physics, Columbia University, New York, NY 10027, USA}

\author{H. Ritsch}
\affiliation{Institute for Theoretical Physics, Innsbruck University, Technikerstrasse 21a, 6020 Innsbruck, Austria}

\date{\today}

\begin{abstract}
Subradiant excited states in finite chains of two-level quantum emitters coupled to a one-dimensional reservoir are a resource for superior photon storage and controlled photon manipulation. Typically, states storing multiple excitations exhibit fermionic correlations and are thus characterized by an anti-symmetric wavefunction, which makes them hard to prepare experimentally. Here we identify a class of quasi-localized dark states with up to half of the qubits excited, which appear for lattice constants that are an integer multiple of the guided-mode wavelength. They allow for a high-fidelity preparation and minimally invasive read out in state-of-the-art setups. In particular, we suggest an experimental implementation using a coplanar wave-guide coupled to superconducting transmon qubits on a chip. As free space and intrinsic losses are minimal, virtually perfect dark states can be achieved even for a low number of qubits, enabling fast preparation and manipulation with high fidelity.

\end{abstract}


\maketitle
\textit{Introduction.---}Collective excitation states of ensembles of quantum emitters posses a wealth of surprising physical properties. Typically, the many-body response of these ensembles leads to delocalized excitations that are lost to dissipation at lifetimes that can vary across many different orders of magnitudes. Of particular interest are dark or sub-radiant states whose long life-times can be used to implement extremely efficient quantum memories~\cite{Asenjo-Garcia2017,Moreno-Cardoner2019}, lossless transport of photons~\cite{masson2020_2, Gutierrez-2022}, photon-photon gates~\cite{Moreno-Cardoner2021}, to realize future generations of atomic lattice clocks~\cite{Madjarov_2019,Norcia_2019} and potentially for improved quantum sensing. Recently, applications towards building superior single photon antennas~\cite{Moreno-Cardoner2020} or nanoscopic coherent or non-classical light sources based on dark resonances were proposed~\cite{holzinger2021}.

In most cases, studies and experiments on subradiance focus on manipulating only a single excitation, i.e., they limit their scope to the lowest of the so-called Dicke manifold~\cite{Mirhosseini2019,Manzoni_2018,ballatine2021_2,kornovan2019,holzinger2021,Corzo2019,olmos2020,orell2022,chu2022,he2020,bigorda2022,bettles2016,Paulisch_2016}.
Many-body subradiant states have attracted more interest only recently, but in general the preparation and manipulation of such states remains challenging as they are typically very delocalized. One option is to use more complex atomic emitters with several internal excited states. This allows to store several photons in a dark subspace, but they are tied to multipartite entanglement, which is fragile in general~\cite{Asenjo-Garcia2019,tudela2017,Holzinger_2020_2,Bigorda2022_2,chen2018}. For a chain of qubits coupled to a waveguide, dark states within the two-excitation sector have been classified into fermionic, dimerized or edge states among others~\cite{zhang2019,Albrecht_2019,henriet2019,zhang2019_2,bakkensen,Fang2014,zhong2020,sheremet2021waveguide,poshakinskiy2021,Poshakinskiy2021_2,jenkins2017,ke2019,masson2020,fayard2021,zhong2021}.
Experimental preparation and control of such states remains challenging and only quite recently the two-excitation sector was probed experimentally with superconducting transmon qubits~\cite{Zanner2022}.

\begin{figure}[ht]
\vspace{-2mm}
\includegraphics[width=0.9\columnwidth]{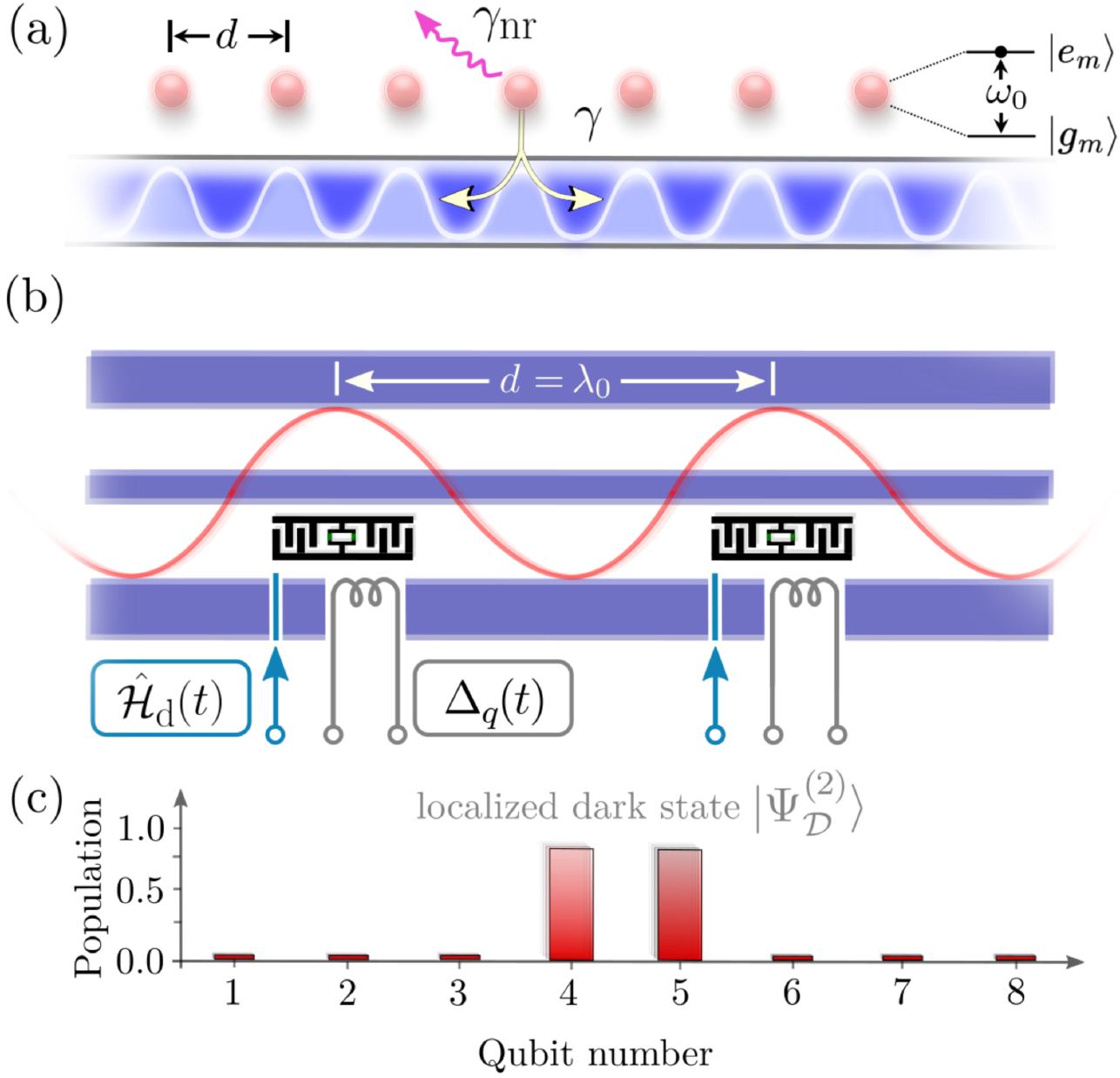}
\vspace{-2mm}
\caption {(a) Schematics of a regular chain of qubits coupled to a 1D waveguide with photon-mediated interactions determined by the single-qubit decay rates $\gamma$. For qubits separated by integer multiples of the wavelength $\lambda_0$, a degenerate family of non-radiative dark states forms, which are only subject to very small free space decay and non-radiative losses $\gamma_\mathrm{nr}$. (b) Waveguide QED realization with superconducting circuits: transmons (in black) are coupled to a coplanar waveguide (in blue). The individual qubit frequencies and thus effectively their distance $d$ can be tuned in-situ via flux-bias lines. For the preparation and read-out of dark states, local driving pulses $\hat{\mathcal{H}}_\mathrm{d}(t)$ and local detuning control $\Delta_\mathrm{q}(t)$ are applied via separate control lines. (c) Distribution of the excited state population for $N=8$ qubits for a typical two-excitation dark state $|\Psi_\mathcal{D}^{(2)}\rangle$ as described by Eq.~\eqref{dark2}. Two qubits store a large fraction $2(N-3)/(N-2)$ of the excitation energy.}
\label{fig1}
\vspace{-8mm}
\end{figure}

\begin{figure*}[ht]
    \begin{center}
    \includegraphics[width=0.97\textwidth]{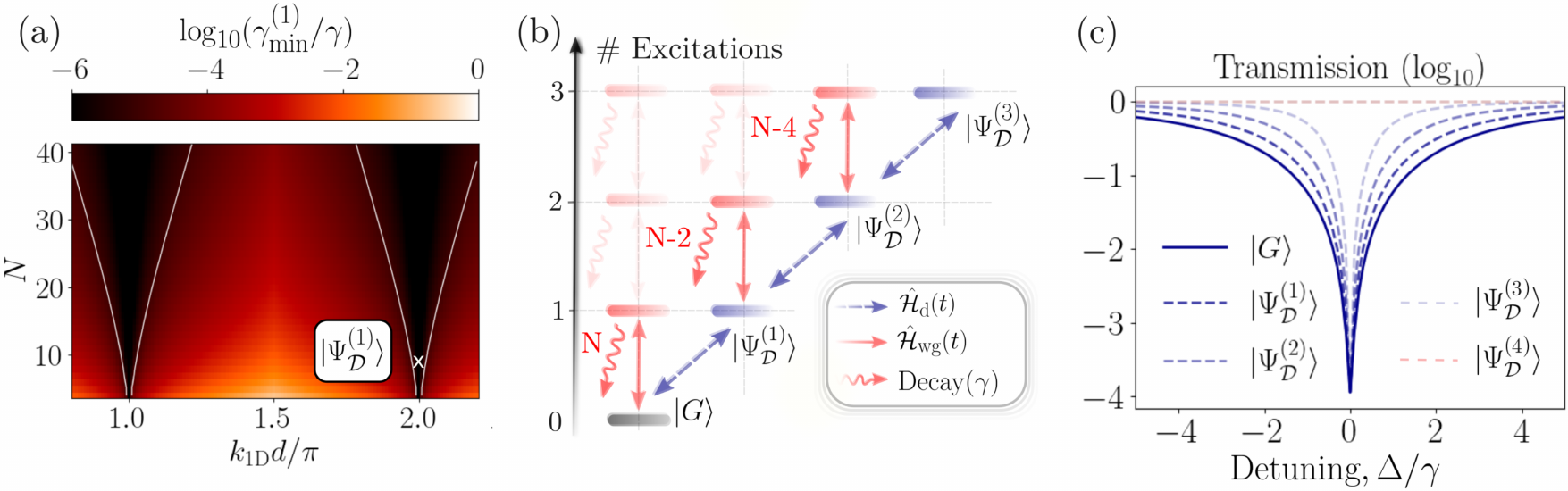} %
     \end{center}
         \vspace{-6mm}
    \caption{(a) Minimal excitation decay rate $\gamma^{(1)}_\mathrm{min}$ within the single-excitation sector as a function of qubit number and separation $d$ for lossless qubits with $\gamma_\mathrm{nr}=0$. The continuous white lines enclose a region of strong collective subradiance, where $\gamma^{(1)}_\mathrm{min}/\gamma \le 10^{-5}$. The example of Eq.~\eqref{dark1} is indicated with a white cross for $N=8$ qubits. (b) Assuming $M$ qubits are driven individually, we show the energy level diagram indicating the route towards dark state preparation and probing with coupling to $|\Psi_\mathcal{D}^{(M)}\rangle$ facilitated by a coherent drive $\hat{\mathcal{H}}_\mathrm{d}(t)$. Once $|\Psi_\mathcal{D}^{(M)}\rangle$ is prepared a second field sent through the waveguide, as described by $\hat{\mathcal{H}}_\mathrm{wg}(t)$ in Eq.~\eqref{waveguide}, transfers the state outside the dark manifold, from where it decays with rate $(N-2M)\gamma$. (c) Weak field waveguide transmission as a function of probe frequency tuned across the single qubit resonance frequency $\omega_0$ for a 8-qubit chain in the ground state (solid line) and the single- to four-excitation dark states (dashed lines). The blockade window decreases from the linewidth $N\gamma$ of the symmetric single-excitation state towards $(N-2M)\gamma$ for the M-excitation dark state and disappears for the four-excitation dark state showing complete transmission.}%
             \vspace{-4mm}
    \label{fig:rabi}%
\end{figure*}

In this work, we theoretically predict a new type of many-body dark states for arrays of qubits coupled to a 1D bath. These states emerge when the lattice constant is an integer of the guided mode wavelength and are distinguished by strongly localized excitations. The states are built from antisymmetric superpositions of symmetric states, whose decay into the bath is forbidden due to destructive interference. For instance, we find that a large fraction $2(N-3)/(N-2)$ of two excitations stored in an $N$ qubit array settles in just two qubits, while a small fraction spreads along the remaining qubits to inhibit decay [see Fig.~\ref{fig1}(c)]. We show below an analytical description for these states and characterize their spatial correlations. We study spectral signatures of photon transport in the presence of these states. From these findings, we propose a realistic protocol to store and release microwave photons in a controlled fashion. Our work should lead to multiple opportunities within atomic physics and quantum optics, such as multi-photon memories for quantum repeaters, and unlock rich phenomena in ordered systems of long-range interacting quantum emitters, both in the linear and quantum many-body regimes. We also note that the high-fidelity preparation protocol presented in this work may inspire experimental confirmation and further the understanding of many-body subradiant states.

\textit{Model.---}Consider an array of $N$ qubits resonantly coupled to the modes of a waveguide as illustrated in Fig.~\ref{fig1}. Each qubit has two internal states $\vert e_{m} \rangle$ and $\vert g_{m} \rangle$ separated by a transition frequency $\omega_{0}$ and is characterized by its position $x_{m}$. The waveguide mediates the qubit-qubit interactions and acts as a source of dissipation. With the inclusion of spontaneous emission into the waveguide and assuming that $\omega_0$ is well below the cutoff frequency of the waveguide, the master ++equation for the density operator of the array $\hat{\rho}$ reads~\cite{lalumi2013,Chang_2012} $\dot{\hat{\rho}} = -i \left(\hat{\mathcal{H}}_\mathrm{eff} \hat{\rho}-\hat{\rho}\hat{\mathcal{H}}_\mathrm{eff}^\dagger\right) + \sum_{m,n} \gamma_{m,n}\hat{\sigma}_m\hat{\rho}\hat{\sigma}_n^\dagger$, where $\hat{\mathcal{H}}_\mathrm{eff}$ is the collective Hamiltonian $(\hbar = 1)$
\begin{equation}
\hat{\mathcal{H}}_\mathrm{eff} = \sum_{m,n=1}^N \Big(J_{m,n}-i\frac{\gamma_{m,n}}{2}\Big)\hat{\sigma}^\dagger_m \hat{\sigma}_{n} \, ,
\label{ham1}
\end{equation}
composed of lowering operators $\hat{\sigma}_m = |g_m \rangle \langle e_m|$ and interaction terms $J_{m,n} = \gamma \sin (k_{0} |x_m-x_n|)/2$ and  $\gamma_{m,n} = \gamma \cos (k_{0} |x_m-x_n|)$. The interaction is weighted by the individual decay rate  $\gamma$ while the qubit separation by $k_{0} =\omega_0/c$, the wavevector of the guided mode on resonance with the qubits. For qubit separation $d=n\lambda_0$ with $n \in \mathbb{N}^+$, the coherent exchange rates $J_{m,n}$ are zero and there is only collective dissipation $\gamma_{m,n}$.

We are interested in localized dark states of many excitations. To construct such states, we divide the chain into two parts of $M$ and $N-M$ qubits, respectively, with $M$ the number of excitations to be manipulated. The precise position of the qubits is not relevant for this division and without losing generality we define the collective operators $\mathcal{S}_1 = \sum_{j=1}^{M} \hat{\sigma}_j/\sqrt{M}$ and $\mathcal{S}_2 = \sum_{j=M+1}^{{N}} \hat{\sigma}_j/\sqrt{N-M}$ to act over each part. The effective Hamiltonian for $d=n\lambda_0$ within this division reads
\begin{align}
\hat{\mathcal{H}}_\mathrm{eff} &=-\frac{iM\gamma}{2}\mathcal{S}_1^{\dagger} \mathcal{S}_1-\frac{i{(N-M)}\gamma}{2}\mathcal{S}_2^{\dagger} \mathcal{S}_2 \nonumber \\
&-i\Gamma(\mathcal{S}_1^{\dagger}\mathcal{S}_2+\mathcal{S}_2^{\dagger}\mathcal{S}_1).
\label{ham2}
\end{align}
The last term shows that the symmetric superpositions of the two parts are dissipatively coupled by the enhanced rate $2\Gamma = \sqrt{M {(N-M)}}\gamma $. There are an additional $N-1$ operators with zero coupling to the waveguide and, therefore, not appearing in Eq.~\eqref{ham2}. Note that the same results follow for odd multiples of $\lambda_0/2$ separations with the symmetric operators now replaced with anti-symmetric operators, having alternate signs between consecutive qubits.

The division is a formal one, but our results can be generalized to non-identical couplings as shown in the in the Supplementary Information~\cite{supp}. In particular, if we assume that the first qubits coupled with a rate $\gamma_{1}$ while the remaining with a rate $\gamma_{2}$ the localization is enhanced. That is, a higher fraction of the excited state population is concentrated in the first qubits. The effects of impurities as non-radiative enegy loss $\gamma_{\text{nr}}$ and dephasing $\gamma_{\phi}$ are also explored in the SI.

\indent \textit{Single Excitation.---}The qubits decay into the waveguide via collective channels determined by the eigenstates of Eq.~\eqref{ham1}. For large arrays, the decay rates depend on qubit number $N$ and lattice spacing $d$, as shown in Fig.~\ref{fig:rabi}(a) for the slowest decay rate. While in general the decay rate of the most subradiant eigenstate is suppressed with increasing qubit number -- following a $N^{-3}$ scaling~\cite{Albrecht_2019} -- this is not the case for spacing $k_\mathrm{1D}d= n \pi$. In this so-called ``mirror configuration'' (with $d=n\lambda_0$), there is only one bright state,  $|\Psi^{(1)}_\mathcal{S}\rangle= \textstyle \sum_j^N \hat{\sigma}_j^\dagger |G\rangle/\sqrt{N}$ where $|G\rangle = |g\rangle^{\otimes N}$, and $(N-1)$ perfectly dark states of exactly zero decay rate.
Leveraging the degeneracy of the dark manifold, one can build highly-localized dark states. Consider the state
\begin{equation}
|\Psi^{(1)}_\mathcal{D} \rangle =\frac{1}{\sqrt{N}}\Big( {\sqrt{N-1}\hat{\sigma}_1^\dagger - {\mathcal{S}}_2^\dagger} \Big) |G\rangle ,
\label{dark1}
\end{equation}
composed of the normalized sum of $| \Psi_m \rangle = 1/\sqrt{2}(\hat{\sigma}^\dagger_1-\hat{\sigma}^\dagger_m)|G\rangle$ states spanning the $N-1$ dark subspace and where ${\mathcal{S}}_2$ is defined in the model section. It displays the unique feature that a large fraction $\langle \hat{\sigma}^\dagger_1 \hat{\sigma}_1\rangle = 1-1/N$ of the excited state population is concentrated in the first qubit. By increasing the system size, the excitation is mostly stored in the first qubit while being protected from decay by a vanishing amount spread in the remaining qubits. The absence of coherent exchange interaction is crucial in so far as it would introduce unwanted couplings between bright and dark states.

\indent \textit{Dark state preparation and probing.---} The choice of storing the excitation in the first qubit is not unique and any other qubit is equally valid~\cite{supp}. The dark state $|\Psi_\mathcal{D}^{(1)}\rangle$, however, can be efficiently prepared by introducing an external coherent drive of frequency $\omega_l$ localized on the first qubit. This pulsed drive on resonance with the first qubit couples to the chain via $\hat{\mathcal{H}}_\mathrm{d}(t)= \Omega_\mathrm{d}(t) (\hat{\sigma}_1^\dagger +\hat{\sigma}_1)$ where $\Omega_\mathrm{d}(t)$ is a time-dependent Rabi frequency. It connects the ground state to both bright and dark states with asymmetrical coupling strengths
\begin{align}
&\langle \Psi_\mathcal{S}^{(1)} | \hat{\mathcal{H}}_\mathrm{d}(t) | G \rangle =  \Omega_\mathrm{d}(t) \sqrt{1/N}, \\
&\langle \Psi_\mathcal{D}^{(1)} | \hat{\mathcal{H}}_\mathrm{d}(t) | G \rangle = \Omega_\mathrm{d}(t){\sqrt{1-1/N}},
\end{align}
preparing the dark state with high fidelity in the $N\gg 1$ limit. The drive not only prepares single-excitation dark states but also connects dark states along the excitation ladder through paths illustrated in Fig.~\ref{fig:rabi}(b). These paths continue until half of the qubits are excited and there are no more dark states~\cite{poshakinskiy2021}.
\begin{figure}[ht!]
    \begin{center}
    {\includegraphics[width=0.97\columnwidth]{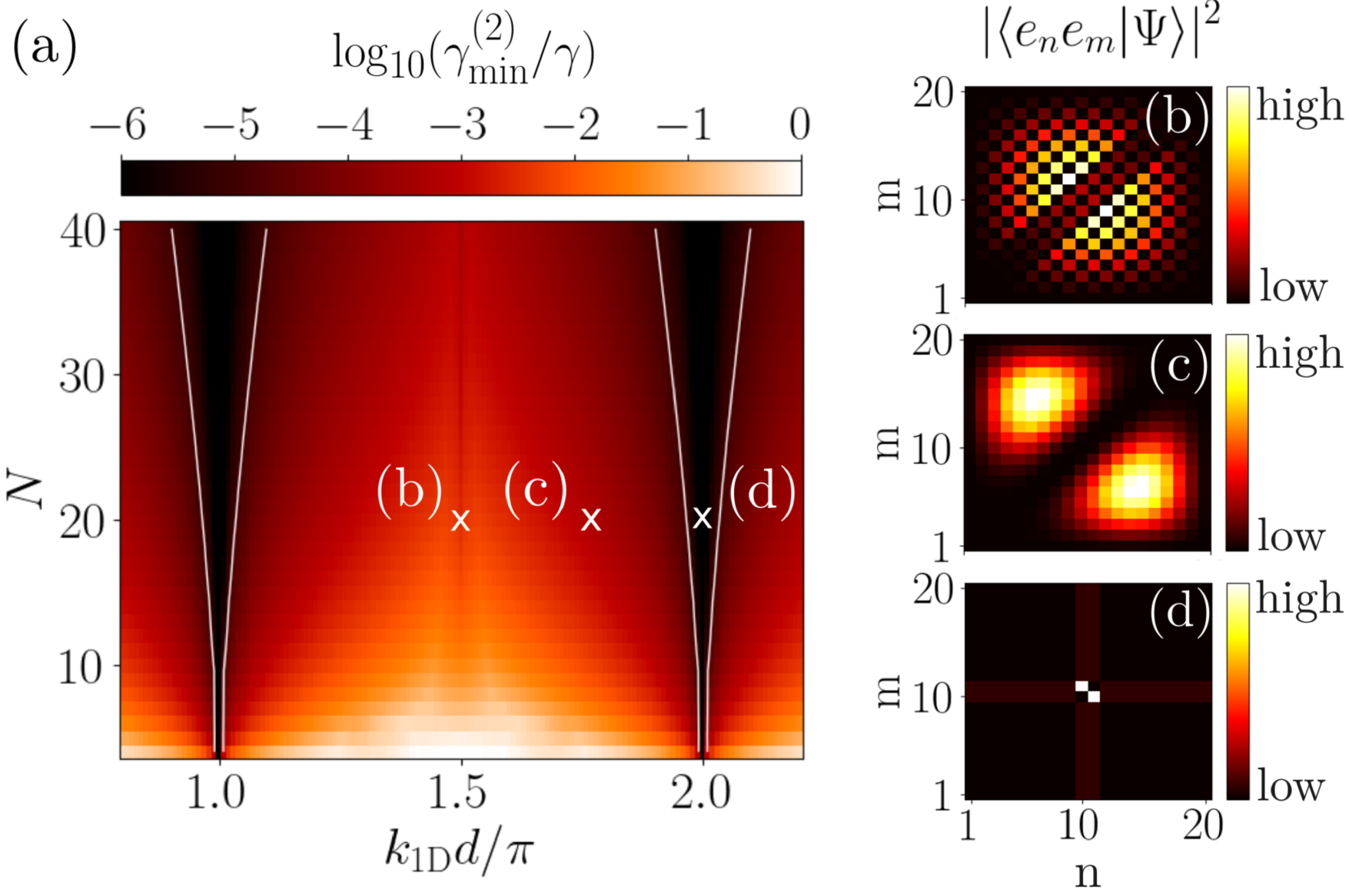} }%
     \end{center}
    \vspace*{-5mm}
    \caption{(a) Minimal decay rate $\gamma_\mathrm{min}^{(2)}$ within the second excitation manifold as a function of chain size $N$ and qubit separation $d$. The continuous white lines enclose the regions where $\gamma^{(2)}_\mathrm{min}/\gamma\le 10^{-5}$. The subradiant states generally exhibit non trivial spatial correlations $|\langle e_n e_m | \Psi \rangle |^2$, which renders them challenging to access. For $k_\mathrm{1D} d = (2n+1)\pi/2$ with $n \in \mathbb{N}$ a checkerboard pattern emerges in (b) whereas in (c) a typical fermionic occupation is shown, which is shared by most subradiant states. In (d) the dark state $|\Psi_\mathcal{D}^{(2)} \rangle$, of Eq.~\eqref{dark2}, is shown for $k_\mathrm{1D}d=2n\pi$ with two excitations localized in the center of the array.}%
    \label{fig:second-ex}%
\end{figure}

To probe the dark states we use a second, weak driving field ($\Omega_\mathrm{wg}(t)/\gamma \ll 1$). The field propagates along the waveguide and couples to the qubits through
\begin{align}
\hat{\mathcal{H}}_\mathrm{wg}(t)= \sum_{j=1}^N \Big( \Delta_\mathrm{wg}\hat{\sigma}_j^\dagger \hat{\sigma}_j + \Omega_\mathrm{wg}(t)(\hat{\sigma}_j^\dagger+\hat{\sigma}_j) \Big) \,.
\label{waveguide}
\end{align}
Notice there is no phase pick-up between the qubits due to the $n\lambda_0$ separation. This probe connects dark and bright states through paths shown in  Fig.~\ref{fig:rabi}(b). It then opens a window into the dark states by measuring the field  $\hat{E}=\hat{E}_\mathrm{in}+i\sqrt{\gamma/2} \sum_j \hat{\sigma}_j$  composed from the superposition of probe and fields scattered into the waveguide.
Figure~\ref{fig:rabi}(c) shows the transmission $\langle \hat{E}^\dagger \hat{E} \rangle / \langle \hat{E}_\mathrm{in}^\dagger \hat{E}_\mathrm{in} \rangle$ for different initial states. An 8-qubit chain is probed by a rectangular waveguide pulse of duration $t \gamma=50$ during which the transmitted field is recorded using the master equation accounting for multiple excitations~\cite{Mirhosseini2019}. Note that we assume the ideal case without imperfections and positional disorder which, would lead to a finite lifetime of the dark state and a higher overall transmission, treated in the SI~\cite{supp}.
We begin with $N$ qubits in the ground state where the transmission linewidth is $N\gamma$, corresponding to the symmetric state $|\Psi_\mathcal{S}^{(1)}\rangle$ excited by the probe. For the qubits prepared in the $M$th excitation dark state the transmission linewidth is reduced to $(N-2M)\gamma$~\cite{supp}. For a single excitation, with $N\ge 3$, the probe excites $\hat{\mathcal{H}}_\mathrm{wg}(t)|\Psi_\mathcal{D}^{(1)}\rangle \propto |\Psi^{(2)}\rangle$, with $|\Psi^{(2)}\rangle \propto ((N-2)\sigma^\dagger_1 -\sqrt{N-1}\mathcal{S}^\dagger_2)\ \mathcal{S}^\dagger_2 |G\rangle$. For $M=N/2$ the waveguide drive is orthogonal to the dark state and therefore renders the system completely transparent. In this way the two-excitation manifold is utilized to escape the decoherence-free subspace and probe the preparation of the dark state~\cite{Zanner2022}.

\indent \textit{Multiple Excitations.---} The localized dark states for multiple excitations are written explicitly in the SI~\cite{supp}. For simplicity, we focus on the two-excitation subspace of Eq.~\eqref{ham2}, where the Hilbert space is spanned by states $|e_n e_m \rangle = \hat{\sigma}^\dagger_n \hat{\sigma}^\dagger_m |G\rangle$ and the most superradiant two-excitation state can be written as $| \Psi_\mathcal{S}^{(2)} \rangle \propto
\sum_{j<k} \hat{\sigma}^\dagger_j \hat{\sigma}^\dagger_k |G\rangle$ with decay rate $2(N-1)\gamma$. The extension of the dark state in Eq.~\eqref{dark1} with $M=2$ (see Eq.~\eqref{ham2}) is
\begin{align}
&| \Psi_\mathcal{D}^{(2)} \rangle = {\frac{\sqrt{N-3}}{\sqrt{N-1}}} \Bigg((\mathcal{S}_1^\dagger)^2 -{\frac{{\sqrt{2} \mathcal{S}_1^\dagger \mathcal{S}_2^\dagger}}{{\sqrt{N-2}}}} + \frac{(\mathcal{S}_2^\dagger)^2}{N-3}   \Bigg)|G \rangle,
\label{dark2}
\end{align}
with $2(N-3)/(N-2)$ of the two excitations stored in the first two qubits.

The advantage and special nature of the ``mirror configuration'' (lattice constant close or equal to $n\lambda_0$) is illustrated in Fig.~\ref{fig:second-ex} where (a) shows the minimal decay rate $\Gamma^{(2)}_\mathrm{min}$ as a function of qubit number $N$ and relative distance $d$. Qualitatively different spatial correlations $|\langle e_n e_m | \Psi \rangle |^2$ of different types of dark states are shown in Figs.~\ref{fig:second-ex}(b)-(c). For $k_\mathrm{1D}d= (2n+1)\pi/2$ with $n \in \mathbb{N}$, the correlations display a checkerboard-type pattern~\cite{Albrecht_2019} due to the fact that the coherent nearest-neighbor and dissipative next-nearest-neighbour interactions in Eq.~\eqref{ham1} are zero. Figure~\ref{fig:second-ex}(c) shows a typical state, described by a fermionic Ansatz, where two-excitation states are composed of single-excitation subradiant states, commonly found as well for multiple excitations~\cite{Asenjo-Garcia2017,zhang2019}. For a large number of qubits, $N \gtrsim 50$ and $k_\mathrm{1D}d= (6n-1)\pi/6$, another extremely subradiant two-excitation state emerges with dimerized spatial correlations and a decay rate lower than any fermionic-type state \cite{zhang2019_2}. In Fig.~\ref{fig:second-ex}(d), the dark state $|\Psi_\mathcal{D}^{(2)}\rangle$ with $k_\mathrm{1D}d=2\pi n$ is shown for a 20-qubit chain and the exact form is depicted in Eq.~\eqref{dark2}.

The spatial correlations of the dark state lead to easily accessible preparation as opposed to most other subradiant states with non-trivial spatial correlations. For instance, a (local) coherent drive with Rabi frequency $\Omega_\mathrm{d}(t)$ exciting two of the qubits drives the dark state with strength $\Omega_\mathrm{d}(t)\sqrt{N-3}/\sqrt{N-1}$ and subsequently a waveguide drive can be used to probe the preparation of the dark state, see also Fig.~\ref{fig:rabi}(b). The transmission in Fig.~\ref{fig:rabi}(c) shows a reduction in linewidth if the dark state $|\Psi_\mathcal{D}^{(M)}\rangle$ is prepared compared to $N\gamma$ if all qubits are in the ground state, due to the fact that the waveguide drive in Eq.~\eqref{waveguide} excites the dark state to $M+1$ excitation states exhibiting decay rates $(N-2M)\gamma$ into the waveguide~\cite{supp}. As a consequence for $N$ being odd or even, the $N/2$ or $(N-1)/2$ excitation dark state leads to unit transmission.

\begin{figure}[ht!]
    \begin{center}
    {\includegraphics[width=0.9\columnwidth]{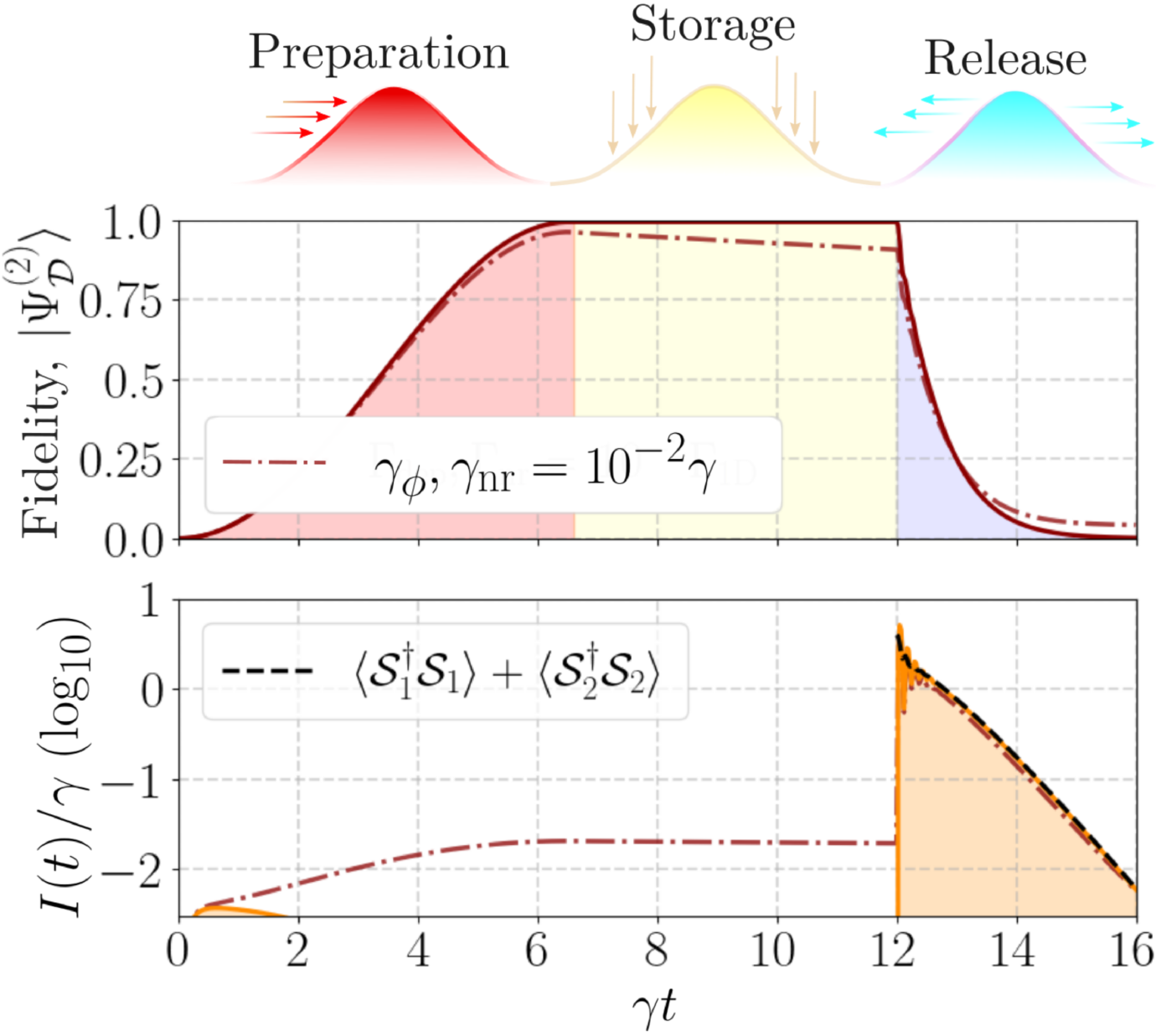} }%
     \end{center}
    \vspace*{-6mm}
    \caption{Protocol to prepare, store, and release two excitations using a chain of $16$-qubits separated a distance $d=\lambda_0$. (a) A $\pi$-pulse drives the first two qubits into the dark state $|\Psi_\mathcal{D}^{(2)}\rangle$ where excitations are stored until $\gamma t= 12$, when they are released via a superradiant channel created by quickly detuning the last $N-2$ qubits by $\Delta_\mathrm{q}= 50\gamma$. (b) Fidelity $F = \langle \Psi_\mathcal{D}^{(2)} \vert \rho \vert \Psi_\mathcal{D}^{(2)}\rangle $ to prepare the dark state for an ideal case (solid line) compared to a case with dephasing and non-radiative damping  $\gamma_{\text{dep}},\gamma_{\text{nr}}=10^{-2}\gamma$ (dashed-dotted). (c) The field radiated into the waveguide displays a sharp peak in intensity $I(t) = \langle \hat{E}^\dagger \hat{E} \rangle(t)$ after release and negligible values during preparation and storage. A beating in intesity appears as the excitation oscillates between initial and final qubits during release  [see $\hat{\mathcal{S}}_{1,2}$ in Eq.~(\ref{ham2})]. Emission with (red dashed) and without interference term $2\mathrm{Re}\langle \mathcal{S}_1^\dagger \mathcal{S}_2 \rangle$ (black dashed). Here, the $\pi$-pulse has a Gaussian temporal profile of duration $8\gamma$ at FWHM and reaches a peak Rabi frequency $0.25\gamma$ at $t_0 = 3\gamma^{-1}$.}%
    \label{fig:release}%
    \vspace{-3mm}
\end{figure}

\indent \textit{Two-Photon Storage and Release.---}Building on the above results we establish a simple protocol for storing and releasing two excitations into the waveguide. Starting with $N$ qubits in the ground state a coherent pulse on the first two qubits prepares the dark state $|\Psi_\mathcal{D}^{(2)}\rangle$. The two excitations can now be stored for a time ${\tau}$ after which the last $N-2$ qubits are detuned by $\Delta_\mathrm{q} \gtrsim (N-2)\gamma$, which transfers most of the two excitations to the product state $|e_1 e_2\rangle$. This is illustrated in Fig.~\ref{fig:release} for a 16-qubit chain where a coherent drive $\hat{\mathcal{H}_\mathrm{d}}(t)=\Omega_\mathrm{d}(t)(\hat{\sigma}_1^\dagger+\hat{\sigma}_2^\dagger+h.c.)$ prepares the state $|\Psi_\mathcal{D}^{(2)}\rangle$ with and without imperfections. At $\gamma t=12$ the last $14$ qubits are detuned by $\Delta_\mathrm{q}=50\gamma$ from the resonance frequency $\omega_0$ which initiates decay of the excitations.
The radiated intensity $I(t) = \langle \hat{E}^\dagger \hat{E} \rangle(t)$, equivalently expressed as $\langle \mathcal{S}_1^\dagger \mathcal{S}_1 \rangle+\langle \mathcal{S}_2^\dagger \mathcal{S}_2 \rangle+2\mathrm{Re}\langle \mathcal{S}_1^\dagger \mathcal{S}_2 \rangle$ is shown as well with a sharp pulse of emission appearing after the detuning is turned on.

\textit{Superconducting circuit implementation.---}Due to near-perfect mode matching, superconducting qubits in a 1D transmission line~\cite{Astafiev2010,THD2020, Brehm2022, Mirhosseini2019} are an ideal platform for realizing this work. Here, we focus on the implementation with transmon qubits capacitively coupled to a common coplanar waveguide as shown schematically in Fig.~\ref{fig1}(c). Similar to Ref.~\cite{VanLoo2013}, the distance $d$ between the qubits on chip is fixed but changing the frequency at which the transmon qubits emit effectively changes their separation. This ensures that we can satisfy $d \sim \lambda_0$, as well as tune qubits on and off resonance via on-chip flux lines. Weakly-coupled control lines realize the drive $\hat{\mathcal{H}}_\mathrm{d}(t)$ and allow to selectively excite the single qubits respectivley in-situ, and thus prepare dark states~\cite{Zanner2022}. Non-radiative decay rates $\gamma_\mathrm{nr}$ and dephasing rates $\gamma_\phi$ for superconducting qubits are usually multiple orders of magnitude smaller than typical couplings to the waveguide $\gamma$, see the SI~\cite{supp}. The achievable parameters are easily sufficient to realize the protocol demonstrated in Fig.~\ref{fig:release} with $\sim 99 \%$ fidelity for the dark state preparation.

\textit{Conclusions.---} Motivated by state of the art implementations of waveguide-coupled superconducting qubits, we introduced and studied a theoretical model of the properties and excitation pathways of multi-excitation dark states. Due to the symmetry and (practically infinite-range) all to all coupling, such system possesses almost degenereate manifolds of multi-excitation states radiatively decoupled from the waveguide if the qubits are positioned at wavelength distance. These states allow to absorb and store multiple photons simultaneously~\cite{supp}, while localizing the majority of the excitation energy in just a handful of qubits. This contrasts with typical free space subradiant states, where each excitation is maximally delocalized. Their localized nature facilitates the preparation of these states via local addressing of individual qubits, which is currently available in state of the art implementations. The system and the proposed protocol also allows for controlled storage and release of multiple photons into the waveguide, pointing towards possible applications for non-classical multi-photon sources or a tailored memory for a quantum repeater. As the projected numbers for experimental realizations seem favorable, we expect to inspire efforts in various quantum simulation platforms including superconducting circuits or Rydberg arrays~\cite{Kuznetsova2016,Norcia_2019}. Similarly, optical waveguide systems~\cite{Corzo2019} and atoms, which are tweezer trapped in optical resonators~\cite{Sauerwein2022}, can be envisaged as an alternative setup.

R.H. and H.R. acknowledge funding from the Austrian Science Fund (FWF) doctoral college DK-ALM W1259-N27 and the FET OPEN Network Cryst3 funded by the European Union (EU) via Horizon 2020. T.~H-D. acknowledges financial support from the Lise Meitner programme of the Austrian Science Fund (FWF), project M3347.  AAG gratefully acknowledges support from the Air Force Office of Scientific Research through their Young Investigator Prize (grant No.~21RT0751), the National Science Foundation through their CAREER Award (No. 2047380), the A. P. Sloan foundation, and the David and Lucile Packard foundation. G.K. acknowledges funding by the European Research Council (ERC) under the European Union's Horizon 2020 research and innovation program (714235).

\onecolumngrid

\appendix


\section{M-Excitation Dark State}
\label{AppendixA}

Given a linear chain of N qubits at multiples of $\lambda_0$ separation coupled to a 1D waveguide we show the generalization of the dark state presented in the main text to M excitations, given $M\le N/2$.
Assuming identical waveguide couplings $\gamma$ for all qubits, the single- and two-excitation dark states can be extended to arbitrary excitations by using the condition that it has to be a combination of the symmetric operators $\mathcal{S}^\dagger_1 = \sum_{j=1}^M \hat{\sigma}^\dagger_j /\sqrt{M}$ and $\mathcal{S}^\dagger_2 = \sum_{j=M+1}^N \hat{\sigma}^\dagger_j /\sqrt{N-M}$, namely $c_k (\mathcal{S}^\dagger_1)^{M-k} (\mathcal{S}^\dagger_2)^{k}$, with $k\in \{0,...,M\}$. Consequently the eigenvalue equation $\hat{\mathcal{H}}_\mathrm{eff} |\Psi_\mathcal{D}^{(M)}\rangle = 0 |\Psi_\mathcal{D}^{(M)}\rangle$ leads to a system of $M+1$ equations, which can be solved. Alternatively the Gram-Schmidt procedure can be applied with the initial eigenstate being the symmetric $M-$excitation eigenstate up to normalization $|\Psi_\mathcal{S}^{(M)}\rangle \propto (\sqrt{M}\mathcal{S}^\dagger_1+\sqrt{N-M}\mathcal{S}_2^\dagger)^M |G\rangle$ with decay rate $\gamma_\mathcal{S} = M(N-M+1)\gamma$. The general expression for the $M-$excitation dark state reads

\begin{align}
\Big| \Psi_\mathcal{D}^{(M)} \Big\rangle
&=\sqrt{\frac{(N-2M+1)!(N-2M)!}{(N-M+1)!(N-M)!}}
\sum_{k=0}^{M}(-1)^k \binom{N-M-k}{M-k}
\Big[\sqrt{M}\mathcal{S}_1^\dagger\Big]^{M-k}\Big[\sqrt{N-M}\mathcal{S}_2^\dagger \Big]^{k} \ \Big|G \Big\rangle.
\end{align}

This is the unique dark state which involves the operators $\mathcal{S}_1$, $\mathcal{S}_2$ and is orthogonal to $|\Psi^{(M)}_\mathcal{S}\rangle$.
The cases for $M=1,2$ are already shown in the main text, the dark state for $M=3$ excitations is given by

\begin{align}
\Big| \Psi_\mathcal{D}^{(3)} \Big\rangle = \sqrt{\frac{N-5}{N-2}} \Bigg[\frac{\sqrt{3}(\mathcal{S}_1^\dagger)^3}{2}-\frac{3(\mathcal{S}_1^\dagger)^2\mathcal{S}_2^\dagger}{2\sqrt{N-3}}+\frac{\sqrt{3}\mathcal{S}_1^\dagger(\mathcal{S}_2^\dagger)^2}{N-4} - \frac{\sqrt{N-3}(\mathcal{S}_2^\dagger)^3}{(N-4)(N-5)} \Bigg] \Big|G \Big\rangle.
\end{align}

Given the expression for the M-excitation dark state, the excited state population in the first M qubits is expressed as
\begin{align}
 \sum_{j=1}^M \langle \Psi_\mathcal{D}^{(M)} | \hat{\sigma}^\dagger_j \hat{\sigma}_j | \Psi_\mathcal{D}^{(M)} \rangle = \frac{N-2M+1}{N-2M+2}.
\label{pop-frac111}
\end{align}
The choice of letting the first M qubits be excited is arbitrary as, for a qubit separation of multiples of $\lambda_0$, any M of the N qubits can be excited in order to prepare the dark state $| \Psi_\mathcal{D}^{(M)} \rangle $.
Also, by letting the distace be odd multiples of $\lambda_0/2$, the same results as above hold, in which case the symmetric operators are replaced by anti-symmetric operators with alternating sign between consecutive qubits.
One more note on the orthogonality of the single-excitation dark state $|\Psi_\mathcal{D}^{(1)}\rangle$ of the main text: let us define another equally valid dark state $|\tilde{\Psi}_\mathcal{D}^{(1)}\rangle=1/\sqrt{N}(\sqrt{N-1}\hat{\sigma}^\dagger_N - \mathcal{S}_2^\dagger)|G\rangle$, where $\mathcal{S}_2^\dagger$ creates the symmetric superposition for the first $N-1$ qubits. It follows that these two dark states are nearly orthogonal for large $N$, namely, $|\langle \tilde{\Psi}_\mathcal{D}^{(1)}|{\Psi}_\mathcal{D}^{(1)}\rangle|=1/(N-1)$.

\section{Bright/Dark Subspaces}
\label{AppendixB}

\begin{wrapfigure}{r}{90mm}
    \vspace{-5mm}
  \begin{center}
    {\includegraphics[width=0.45\columnwidth]{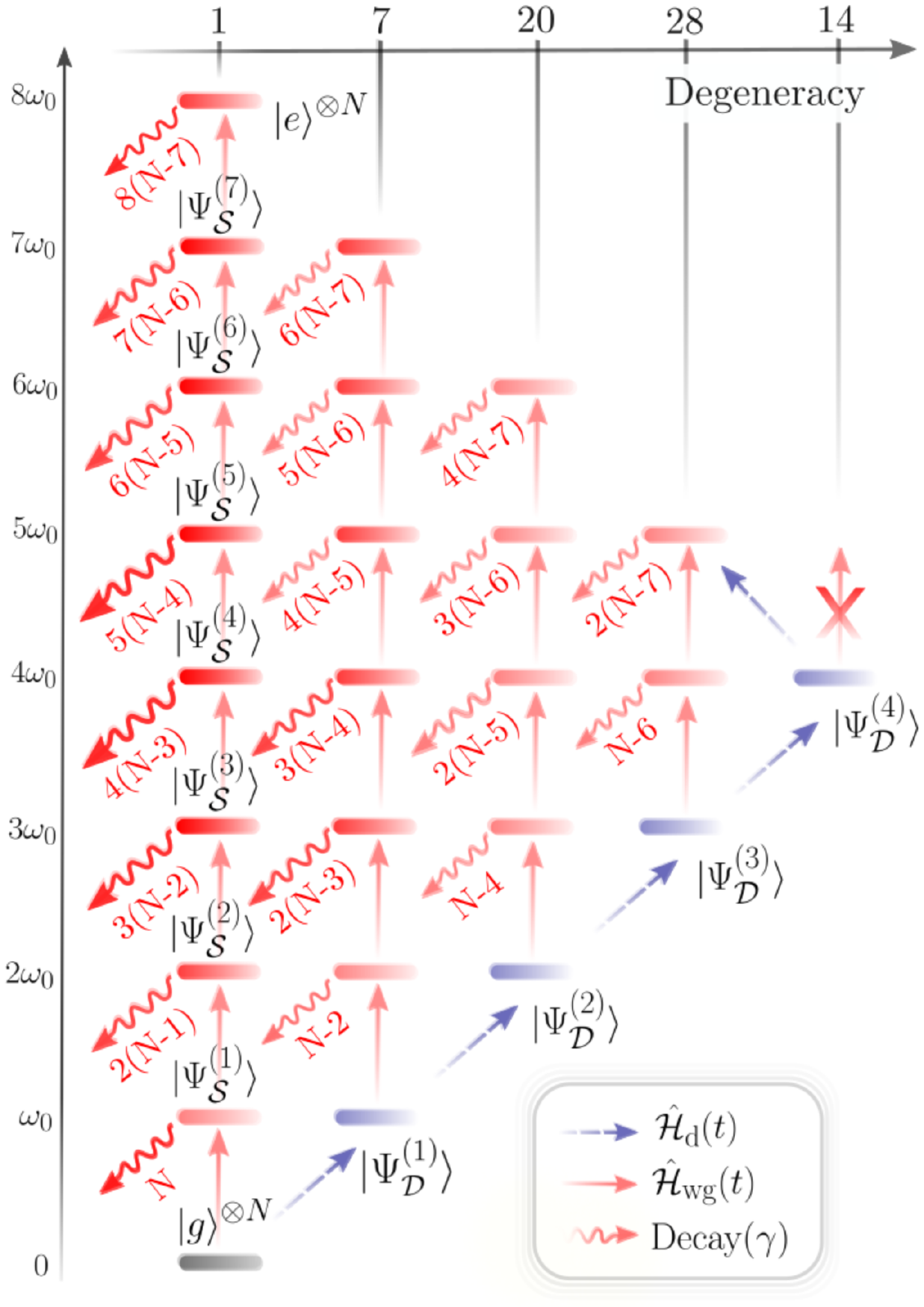} }%
  \end{center}
      \vspace{-0mm}
  \caption{Energy level diagram for $N=8$ qubits with multiples of $\lambda_0$ separation featuring a coherent drive $\hat{\mathcal{H}}_\mathrm{d}(t)$ on individual qubits and a waveguide drive $\hat{\mathcal{H}}_\mathrm{wg}(t)$ exciting all qubits symmetrically. The symmetric Dicke states $|\Psi_\mathcal{S}^{(M)}\rangle$ decay with the superradiant rate $M(N-M+1)\gamma$ which a maximum at $M=N/2$. Dark states exist for $M\le N/2$ and $|\Psi_\mathcal{D}^{(M)}\rangle$, which resides in the dark subspace with $M$ excitations, can be efficiently prepared by driving $M$ qubits. The state can be probed via a symmetric waveguide drive on all qubits which excites it into a bright manifold with decay rate $(N-2M)\gamma$. Qubits prepared in the state $|\Psi_\mathcal{D}^{(N/2)}\rangle$ lead to complete transmission, as there is no bright manifold above with the same $s$ quantum number. In between symmetric Dicke states and the dark state manifold are bright state manifolds with finite decay rates which are connected by a symmetric excitation to the excitation manifold below. Each subspace has a certain degeneracy which is shown on the top and note that there is no mixing between states from different excitation manifolds due to the absence of collective dephasing (no imperfections).}%
    \label{dark-manifold}%
    \vspace{-46mm}
\end{wrapfigure}

In this section we explain the possible excitation/decay paths for the highly degenerate effective Hamiltonian at multiples of $\lambda_0$ qubit separation in more detail, in particular to understand the transmission properties if a certain dark state is prepared. Using the language of Dicke superradiance for collective spins, the symmetric collapse operator for all qubits, $\mathcal{S} = (\sqrt{M}\mathcal{S}_1 + \sqrt{N-M}\mathcal{S}_2)/\sqrt{N}$ is in the angular momentum representation for the sum of $N$ spin $1/2$ subsystems and defines lowering/raising operations on the Dicke states, $\mathcal{S}|N/2,m\rangle \propto |N/2,m-1\rangle$ and $\mathcal{S}^\dagger|N/2,m\rangle \propto |N/2,m+1\rangle$ where states are generally expressed as $|m,s\rangle$ with the quantum numbers $s$ running from 0 or 1/2 to $N/2$ and $m$ from $-s$ to $s$. The Dicke states are explicitly given by the symmetric M-excitation states $|\Psi_\mathcal{S}^{(M)}\rangle$ with a decay rates $M(N-M+1)\gamma$ and are shown in Fig. \ref{dark-manifold}. There are additional bright states, for instance $|N/2-1,m\rangle$ with a $(N-1)$-fold degeneracy and decay rates $(M-1)(N-M)\gamma$.
For multiples of $\lambda_0$ separation, the Dicke limit, jumps between different excitation manifolds can only be achieved with the symmetric operator $\mathcal{S}$ and a waveguide drive on resonance with the qubit transition frequency $\omega_0$ that excites all qubits equally can be expressed as
\begin{align}
\hat{\mathcal{H}}_\mathrm{wg}(t) = \Omega_\mathrm{wg}(t)\Big(\sqrt{M}\mathcal{S}_1^\dagger+\sqrt{N-M}\mathcal{S}_2^\dagger+h.c.\Big),\label{wg_drive}
\end{align}

where the qubit chain is partitioned into $M$ and $N-M$ qubit arrays respectively, and $\Omega_\mathrm{d}(t)$ is the time dependent Rabi frequency which can be pulsed or continuous. The waveguide drive can now connect Dicke states with the same quantum number $m$ shown as vertical red arrows in Fig. \ref{dark-manifold} but is not able to access the completely dark state manifold which requires some asymmetry that is provided here by a coherent drive on a single or multiple qubits, $\hat{\mathcal{H}}_\mathrm{d}(t)=\Omega_\mathrm{d}(t)\sum_m(\hat{\sigma}^\dagger_m+\hat{\sigma}_m)$, with the requirement that only up to $N-1$ qubits are driven, in order to have non-zero overlap with the dark state manifold.
This way, the every dark state (up to the $N/2$-excitation manifold) can be driven. Alternatively, the single-excitation dark state $|\Psi_\mathcal{D}^{(1)}\rangle$ can be prepared by driving a single qubit, subsequently $|\Psi_\mathcal{D}^{(2)}\rangle$ can be prepared by driving another qubit and so on up to $|\Psi_\mathcal{D}^{(N/2)}\rangle$ as shown in Fig. \ref{dark-manifold}. On the other hand, for instance, the waveguide drive in Eq.~\eqref{wg_drive} drives the single-excitation dark state $|\Psi_\mathcal{D}^{(1)}\rangle$ to the two-excitation bright state
\begin{align}
|\Psi^{(2)} \rangle = \sqrt{\frac{2}{3}} \Big(\hat{\sigma}_1^\dagger \mathcal{S}_2^\dagger-\frac{\sqrt{N-1}}{N-2}(\mathcal{S}_2^\dagger)^2 \Big)|G\rangle,
\end{align}
which decays with a rate $(N-2)\gamma$ and is observable in the transmission spectrum. The symmetric single excitation state $|\Psi_\mathcal{S}^{(1)}\rangle$ is driven to the symmetric two-excitation Dicke state  with the superradiant decay rate $2(N-1)\gamma$. All states shown in Fig. \ref{dark-manifold} except the symmetric Dicke states $|\Psi_\mathcal{S}^{(M)}\rangle$, the ground state $|g\rangle^{\otimes N}$ and the totally inverted state $|e\rangle^{\otimes N}$ are elements of a subspace with degeneracy $d_M = \binom{N}{N-M}-\binom{N}{N-M-1}$.

The driving strength (for instance from the single- to the two-excitation dark state) is given by
\begin{align}
&\langle \Psi^{(2)}_\mathcal{D}|\hat{\mathcal{H}}_\mathrm{d}| \Psi^{(1)}_\mathcal{D} \rangle/\Omega_\mathrm{d}(t) = \nonumber \\ &{\frac{\sqrt{N-3}}{\sqrt{N}}} \Bigg( 1+ {\frac{1}{{\sqrt{(N-1)(N-2)}}}}\Bigg),
\end{align}
where the coherent drive only excites the second qubit and the single-excitation dark state has the majority of the excited state population in the first qubit.

\newpage

\section{Effect of Disorder, Imperfections and Non-Identical Waveguide Couplings}
\label{AppendixC}

The above results can be generalized to non-identical waveguide couplings, in particular we assume the first qubit is coupled with rate $\gamma_\mathrm{1}$ and the remaining qubits with rate $\gamma_\mathrm{2}$. Here we show this actually enhances the effect studied so far, that is, an even higher fraction of the excited state population is concentrated in the first qubit and the driving strength of the dark state is equally enhanced.
The single-excitation symmetric state can now be written as
\begin{equation}
|\Psi^{(1)}_\mathcal{S} \rangle = \frac{1}{\sqrt{\gamma_\mathrm{1}+(N-1)\gamma_\mathrm{2}}} \Bigg(\sqrt{\gamma_\mathrm{1}}\hat{\sigma}_1^\dagger + \sqrt{(N-1)\gamma_\mathrm{2}}\mathcal{S}_2^\dagger \Bigg) \Big|G \Big\rangle,
\end{equation}
where the populations in the individual qubits is not equally distributed since the different couplings lead to a redistribution of the excitation. The dark state obtains is readily found to be
\begin{equation}
|\Psi^{(1)}_\mathcal{D} \rangle = \frac{1}{\sqrt{\gamma_\mathrm{1}+(N-1)\gamma_\mathrm{2}}} \Bigg(\sqrt{(N-1)\gamma_\mathrm{2}} \hat{\sigma}_1^\dagger -  \sqrt{\gamma_\mathrm{1}} \mathcal{S}_2^\dagger \Bigg) \Big|G \Big\rangle.
\end{equation}

\begin{figure}[ht!]
    \begin{center}
    {\includegraphics[width=0.99\textwidth]{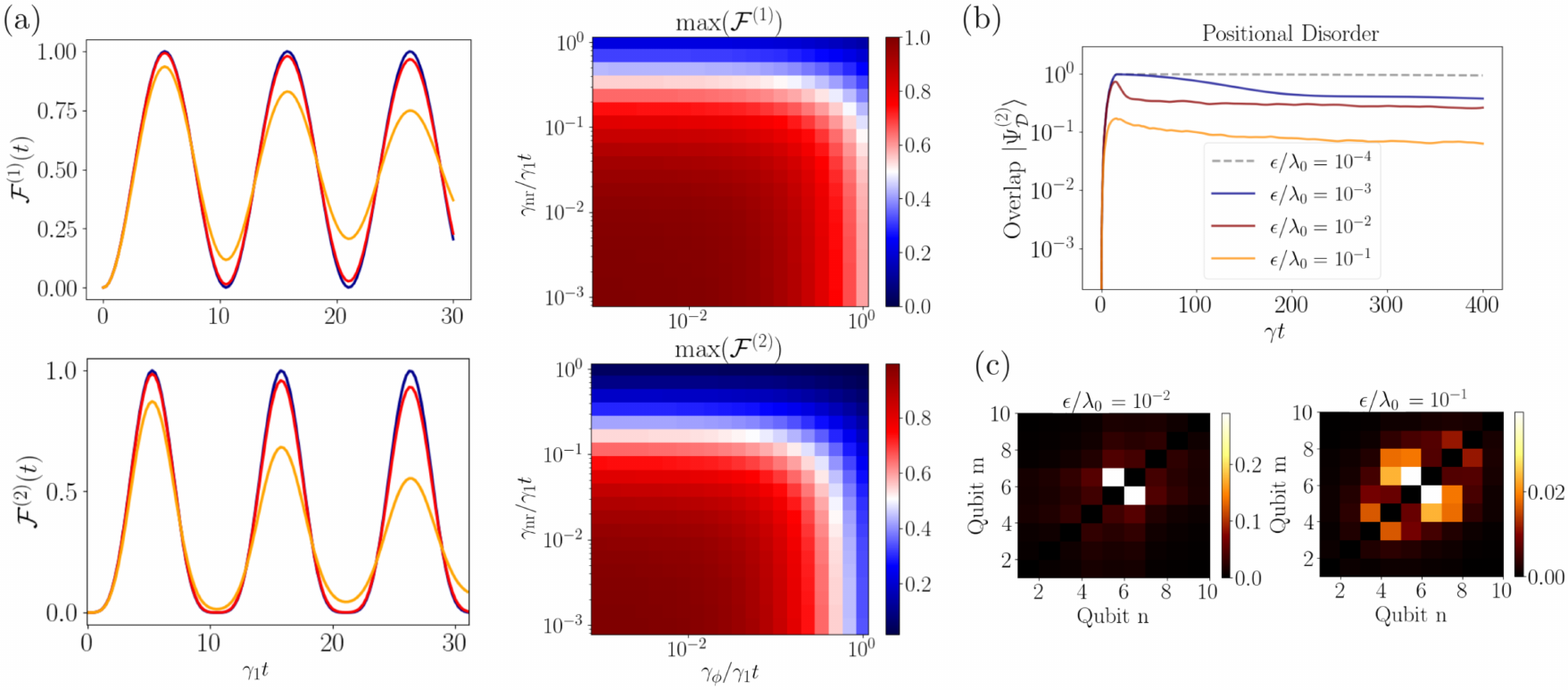} }%
     \end{center}
    \caption{(a) Dark state preparation under dephasing and non-radiative decay for $N=6$ qubits with $\gamma_\mathrm{2}=20\gamma_\mathrm{1}$ for either the last $4$ or last $5$ qubits and $d=\lambda_0$ and under continuous driving with Rabi frequency  $\Omega_\mathrm{d}/\gamma_\mathrm{1}$=0.3 for either the first or the first two qubits. The time evolution of the fidelity with the target state $\mathcal{F}^{(M)}(t) = \langle \Psi_\mathcal{D}^{(M)} | \rho(t) | \Psi_\mathcal{D}^{(M)} \rangle$ is shown. Blue, red and orange lines correspond to dephasing rates (i) $\gamma_\phi=0$, (ii) $\gamma_\phi=0.01 \gamma_\mathrm{1}$, (iii) $\gamma_\phi=0.1 \gamma_\mathrm{1}$ respectively, with $\gamma_\mathrm{nr}$=0. Also shown is the maximal fidelity for preparing the one- or two-excitation dark state $|\Psi_\mathcal{D}^{(M)}\rangle$ in the presence of non-radiative loss and dephasing rates $\gamma_\mathrm{nr}$ and $\gamma_\mathrm{\phi}$. (b) Influence of classical position disorder $\epsilon/\lambda_0$ on the two-excitation dark state preparation with a rectangular driving pulse on the central two qubits and decay rate of the dark state for identical waveguide couplings $\gamma$ in a chain of $N=10$ qubits. The disorder is assumed to be following a normal distribution of standard deviation $\epsilon$ and 200 random configurations for each value of $\epsilon$ are considered. (c) The averaged spatial correlation $|\langle e_n e_m | \Psi(t) \rangle|^2$ is shown at $\gamma t = 20$ for the disordered arrays in (b).}%
    \label{fig:rabi333}%
\end{figure}

The amount of population in the first qubit for the dark state is given by $(N-1) \gamma_\mathrm{2}/(\gamma_\mathrm{1}+(N-1)\gamma_\mathrm{2})$ and in particular for $\gamma_\mathrm{2} \gg \gamma_\mathrm{1}$ most is stored in the first qubit even for small $N$. It follows that the dark state can be prepared even more efficiently given that $\gamma_\mathrm{1}<\gamma_\mathrm{2}$.

The two-excitation dark state, with the first two qubits having a coupling rate $\gamma_\mathrm{1}$, is given by
\begin{align}
&| \Psi_\mathcal{D}^{(2)} \rangle = \frac{1}{\sqrt{\alpha}} \Bigg(\frac{\sqrt{\gamma_\mathrm{1} \gamma_\mathrm{2}}} {\gamma_\mathrm{1}}(\mathcal{S}_1^\dagger)^2 + \frac{\sqrt{\gamma_\mathrm{1}\gamma_\mathrm{2}}}{\gamma_\mathrm{2}(N-2)} (\mathcal{S}_2^\dagger)^2 -{\frac{\sqrt{2}}{\sqrt{N-1}}} \mathcal{S}_1^\dagger \mathcal{S}_2^\dagger  \Bigg)\Big|G \Big\rangle,
\end{align}
with
\begin{align}
\alpha = \frac{\gamma_\mathrm{2}^2 (N-1)(N-2)+2\gamma_\mathrm{1}\gamma_\mathrm{2}(N-2)+2\gamma_\mathrm{1}^2}{\gamma_\mathrm{1}\gamma_\mathrm{2}(N-2)}.
\end{align}

The case of a single qubit interacting with a chain of qubits in the collective symmetric state can also be realized with two-level atoms in free space. In this analogy the first qubit is placed at the center of a ring of subwavelength-spaced two-level emitters where, due to permutational ring symmetry, only the symmetric mode of the ring couples to the central emitter~\cite{Holzinger2020}. The ring plays the role of an antenna focusing most of the incoming radiation into the central emitter, equivalent to driving $N-1$ qubits in the waveguide scenario.

In a realistic scenario both dephasing $\gamma_\phi$ and excitation loss $\gamma_\mathrm{nr}$ into channels other than the waveguide are present and affect the fidelities of driving the one- and two-excitation dark state as well as their lifetimes. The non-waveguide decay and dephasing are included as uncorrelated terms in the Lindbladian and the  master equation for arbitrary distances obtains the form
\begin{align}
\dot{\hat{\rho}} = -i [\hat{\mathcal{H}}_\mathrm{eff} ,\hat{\rho}] + \sum_{m,n} \Gamma_{m,n}\hat{\sigma}_m\hat{\rho}\hat{\sigma}_n^\dagger + \gamma_\mathrm{nr} \sum_{m} \hat{\sigma}_m\hat{\rho}\hat{\sigma}_m^\dagger +2{\gamma_\phi}\sum_{m} \hat{\sigma}^\dagger_m \hat{\sigma}_m \hat{\rho} \hat{\sigma}^\dagger_m \hat{\sigma}_m,
\end{align}
and the effective Hamiltonian in the interaction picture is given by \begin{equation}
\hat{\mathcal{H}}_\mathrm{eff} = \sum_{m,n} \Big(J_{m,n} - i\frac{\Gamma_{m,n}}{2} \Big)\hat{\sigma}^\dagger_m \hat{\sigma}_{n} -i\frac{\gamma_\mathrm{nr}+2\gamma_\phi}{2}\sum_m \hat{\sigma}^\dagger_m \hat{\sigma}_{m}.
\label{ham3}
\end{equation}

For instance in Fig. \ref{fig:rabi333}(a) the influence of individual qubit dephasing and non-radiative decay on the dark state preparation fidelity is shown for non-identical waveguide couplings.
 A high preparation fidelity is prevailing even under considerable individual dephasing and decay, which in state-of-the-art laboratories working with superconducting qubits can be held below $10^{-2}\gamma_\mathrm{1}$ for both. In Fig. \ref{fig:rabi333}(b) the two-excitation dark state preparation and decay rate are shown for various degrees of classical position disorder. Each qubit is randomly displaced around its multiple of $\lambda_0$ position by a normal distribution of standard deviation $\epsilon$. The overlap with $|\Psi_\mathcal{D}^{(2)}\rangle$ is plotted after performing an average over disorder realizations which is confined along the 1D chain axis. The number of disorder realizations is 200 and in Fig. \ref{fig:rabi333}(c) the averaged spatial correlation $|\langle e_n e_m | \Psi(t) \rangle|^2$ is shown after the overlap reached the maximum value. For quantum platforms in the microwave regime e.g. superconducting transmon qubits, the positional disorder can be kept well below $10^{-4}$ but even other platforms like atoms trapped along a nanofiber exhibit small positional disorder.

 Frequency disorder might be a further complication which introduces finite lifetimes to the dark state and decreases the preparation fidelity. Specifically for superconducting circuits this can be remedied by additionally adding flux bias lines, which ensures that we can tune the frequency of all qubits on resonance. Although this means there is an overhead of one control line per qubit, which should be fine for systems up to e.g. 10-15 qubits or potentially even more.
 Another point is that with increasing qubit number $N$, the resilience to frequency disorder increases as, on the one hand, the linewidth of the symmetric state scales with $N$ allowing for more detuned remaining qubits. On the other hand with more qubits it is more likely that at least a handful are closer to resonanance with each other, thereby allowing them to combine into a dark state.

\section{Multilevel Nature of the Transmon Qubit}
\label{AppendixD}

For the specific platform of superconducting transmon qubits the involved quantum emitters are inherently anharmonic and due to the multilevel nature it is necessary to model each emitter as a multilevel quantum emitter to capture the full richness of possible quantum states. So far we assumed a large enough anharmonicity $U$ between the first and second excited state in each transmon and therefore neglected the second excited state to recover the two-level qubit. Here we show that the results obtained above still prevail in the case of small $U$ and even the harmonic oscillator case $U=0$. Let us first begin to write down the effective Hamiltonian for N transmons, which is given by
\begin{equation}
\hat{\mathcal{H}}_\mathrm{eff} = \sum_{m,n} \Big(J_{m,n} - i\frac{\Gamma_{m,n}}{2} \Big)\hat{a}^\dagger_m \hat{a}_{n}-\frac{U}{2}\sum_m \hat{n}_m(\hat{n}_m-\mathbb{I}),
\label{hamD}
\end{equation}
whereas the Lindbladian retains the same form as in the qubit case. The bosonic operator $\hat{a}^\dagger_m$ creates an excitation on the site $m$ and $\hat{n}_j$ is the number operator of the site $m$. For many-body dynamics the anharmonicity U serves as an on-site interaction and the weaker the anharmonicity, the closer the system resembles the harmonic oscillator. For arbitrary $U$ the transmons generally behave like anharmonic oscillators but assuming only a single excitation is present in the system, the single-excitation dark and bright states have the same form as in the qubit case with the exchange $\hat{\sigma}_m \leftrightarrow \hat{a}_m$.
The completely symmetric state with two excitations can be written as
\begin{equation}
|\Phi_\mathcal{S}^{(2)}\rangle = \frac{\sqrt{2}}{N} \sum_{n,m} \hat{a}^\dagger_n \hat{a}^\dagger_m |G\rangle  = \frac{\sqrt{2}}{N} \Big((\hat{a}_1^\dagger)^2 +\sqrt{N-1} \hat{a}_1^\dagger \mathcal{S}_2^\dagger+ (N-1)(\mathcal{S}_2^\dagger)^2 \Big)|G\rangle,
\end{equation}
with the superradiant decay rate $2N\gamma$ which is larger than the superradiant rate of the two-excitation symmetric Dicke state. The two-excitation dark state with two excitations per site and the majority of the excitation in the first transmon is given by
\begin{equation}
|\Phi_\mathcal{D}^{(2)}\rangle = \frac{\sqrt{2}}{N} \Big((\hat{a}_1^\dagger)^2 -2\sqrt{N-1} \hat{a}_1^\dagger \mathcal{S}_2^\dagger+ (\mathcal{S}_2^\dagger)^2 \Big)|G\rangle,
\end{equation}
with the fraction $2(N-1)/N$ of the population in the first transmon and $N\ge 2$. As before, any transmon is equally valid to store the majority of the population and which one is determined whether or not it is excited by an external drive. The other possible dark state involving superpositions of symmetric states in the sub-arrays is equivalent to the two-excitation dark state for the qubit case $|\Psi_\mathcal{D}^{(2)}\rangle$ with the exchange $\hat{\sigma}^\dagger_m \leftrightarrow \hat{a}^\dagger_m$ in $\mathcal{S}_1^\dagger$ and $\mathcal{S}_2^\dagger$ respectively.
Assuming that the first two transmons are driven by an external drive, the general dark state for arbitrary anharmonicities $U$ is a superposition of the aforementioned dark states, namely $c_1 |\Psi_\mathcal{D}^{(2)}\rangle + c_2 |\Phi_\mathcal{D}^{(2)}\rangle$. Notably, the state $|\Phi_\mathcal{D}^{(2)}\rangle$ acquires a energy shift as well coming from the onsite interaction term in the Hamiltonian and has to be taking into account while driving the quibts externally.
This shows that the effect described for the qubit case extends to the transmonic regime with the collective dark and bright states having a slightly modified distribution in the populations due to their multilevel structure.
Typical values for the anharmonicity are $U \sim 200-300 \ \mathrm{MHz}$ whereas the waveguide decay rate $\gamma$ is in the range of $1-100 \ \mathrm{MHz}$.

\end{document}